# On the Origin of Carrier Loss in Mg-Doped N-Polar GaN

Masahiro Kamiyama[1,a], Shashwat Rathkanthiwar[1], Cristyan Quiñones-García[1], Seiji Mita[2], Dolar Khachariya[2], Pramod Reddy[2], Ronny Kirste[2], Ramón Collazo[1], Zlatko Sitar[1,2]

[1]*Department of Materials Science and Engineering, North Carolina State University, Raleigh, North Carolina 27695, USA*
[2]*Adroit Materials, 2054 Kildare Farm Road, Cary, North Carolina, 27518, USA*
a) Author to whom correspondence should be addressed: mkamiya@ncsu.edu

**Abstract**

The neutral $(V_N\text{-}3Mg_{Ga})^0$ complex was found to be the primary compensator in Mg-doped, N-polar GaN. The experimental data showed a sharp drop in hole concentration once [Mg] exceeded $\sim 10^{19}$ cm$^{-3}$. Temperature-dependent Hall measurements, in conjunction with a charge balance model, revealed that the carrier loss was due to a drastic reduction in acceptor concentration ($N_A$), suggesting that a significant fraction of Mg atoms incorporated in an electrically neutral configuration. A quantitative semi-empirical model based on the grand canonical formalism pointed to the formation of $(V_N\text{-}3Mg_{Ga})^0$ complexes as the primary cause for the observed carrier loss.

## I. Introduction

Nitrogen-polar (N-polar) GaN has attracted significant attention for the development of next-generation power and high-frequency devices. The reversed polarization direction of N-polar GaN, as compared to that of Ga-polar GaN, allows for device designs with superior performance. For example, AlGaN/GaN high electron mobility transistors (HEMTs) with the N-polar structure enable a strong back-barrier,[1] low-resistivity Ohmic contacts,[2] and improved scalability.[3] In addition, by taking advantage of the inherent polar doping selectivity, lateral polar junctions realized on side-by-side grown Ga- and N-polar domains allow for the fabrication of novel device structures such as GaN-based superjunctions (SJ),[4,5] lateral p-n homojunctions,[6] and depletion-mode MESFETs.[7] However, the development of bipolar devices has been limited due to the challenges associated with achieving controllable p-doping in N-polar GaN.

Similar to the Ga-polar GaN, there are two inherent challenges in achieving efficient p-doping in N-polar GaN. First, the high Mg acceptor ionization energy (150 to 220 meV) limits the free hole concentration at room temperature (RT) to a few % of the Mg doping concentration.[8–10] Second, the formation of compensating defect, such as nitrogen vacancy-related complex with the Mg acceptor $(V_N\text{-}nMg_{Ga})^{3-n}$, where n = 1, 2 or 3, becomes more likely at higher doping concentrations, limiting the attainable free hole concentrations.[11] There are only limited reports on p-doping of N-polar GaN, likely due to the additional challenges in the growth of this material. For example, without an appropriate surface kinetics control of the growth process, N-polar GaN exhibits a rough surface characterized by hexagonal hillocks[12–14] and incorporates oxygen ($O_N^{+1}$) on the order of $10^{19}$ cm$^{-3}$,[12,15,16] which acts as a compensator for any p-dopant, thereby posing additional challenges in achieving desired p-conductivity.

A "knee" behavior is often reported for Mg-doped Ga-polar GaN as a function of Mg concentration, wherein the hole concentration drops precipitously after a critical Mg doping concentration of about $3\times10^{19}$ cm$^{-3}$ has been reached,[17–22] corresponding to a maximum hole concentration in the range of mid-$10^{17}$ cm$^{-3}$. While this compensation phenomenon is observed universally, its mechanism in p-GaN remains controversial: some attribute it to the formation of pyramidal inversion domains (PIDs), which bind and inactivate Mg dopants,[21,23,24] others argue that the amount of segregated Mg atoms at PIDs is insufficient to explain the knee behavior and point to the presence of compensating point defects and related complexes.[25,26] Several reports have attributed self-compensation to isolated nitrogen vacancies ($V_N^{+3}$) and their complexes with Mg atoms $(V_N\text{-}nMg_{Ga})^{3-n}$, where n = 1, 2, or 3.[17–19] However, the exact nature, i.e., structure and charge state of the responsible defect, has not been studied. The "knee" behavior is typically perceived as the Mg doping limit in GaN for specific growth conditions. Therefore, the identification of the defect responsible for compensation is of significant interest in the quest to achieve higher hole concentrations and improve p-conductivity in GaN via active point defect management techniques,[27–30] particularly when grown with the N-polarity.

In this work, we present a comprehensive study of compensation mechanism in Mg-doped N-polar GaN. First, a charge balance model based on the Fermi Dirac statistics is adopted to extract the acceptor ($N_A$) and donor ($N_D$) concentrations from temperature-dependent Hall effect measurements. Then, a semi-empirical model based on the grand canonical formalism is developed to quantitatively describe and predict the "knee" behavior.

## II. Experimental

All N-polar GaN samples in this study were grown on c-plane sapphire substrates with 4° miscut towards m-plane. The growth was performed in a vertical, cold-walled, low-pressure (20 Torr) metalorganic chemical vapor reactor.[31] Triethylgallium (TEG) and ammonia were used as gallium and nitrogen precursors, respectively. Bis(cyclopentadienyl) magnesium ($Cp_2Mg$) was used as Mg dopant precursor. Magnesium and oxygen concentrations were measured using secondary ion mass spectroscopy (SIMS). To obtain a smooth surface devoid of hillocks, a 160-nm thick surface morphology control layer (SMCL) was first deposited under low supersaturation growth conditions.[12,13] This unintentionally doped layer was found to incorporate oxygen on the order of $10^{19}$ cm$^{-3}$, leading to high n-type conductivity. To minimize any parasitic conduction, this layer was compensated with Mg at a doping concentration of $2\times10^{20}$ cm$^{-3}$, which rendered it highly resistive (~$10^5$ Ω·cm). Following the SMCL, a 1.2-μm thick Mg-doped p-GaN layer was deposited.

To minimize the formation of compensating donors ($N_D$), the p-type layer was grown using high nitrogen chemical potential conditions: $NH_3$ flow rate of 6 slm, TEG flow rate of 133 μmol/min, total flow rate of 7.2 slm, growth temperature of 1313 K, and a pressure of 20 Torr under the $N_2$ diluent.[10,12] The Mg doping concentration was varied from $7x10^{18}$ to $3x10^{19}$ cm$^{-3}$. All samples were annealed in situ under $N_2$ flow for 20 mins at 1173 K to activate the Mg acceptors.[32]

The resulting surface morphology was inspected using an Asylum Research MFP-3D atomic force microscope (AFM) operated in tapping mode. The nitrogen polarity was confirmed using a 1M potassium hydroxide (KOH) etch at 340 K for 1 hour. The threading dislocation density (TDD) was estimated from X-ray diffraction (XRD) measurements using a Phillips X'pert materials research diffractometer.[33]

Ni/Au (20/40 nm) contacts were e-beam deposited in the van der Pauw geometry on 1x1 cm$^2$ samples for electrical measurements. The contacts were annealed at 875 K for 10 minutes in $O_2$ ambient for ohmic contact formation. Temperature-dependent resistivity and Hall measurements were performed in a temperature range of 220-400 K using a Lake Shore 8400 series AC/DC Hall measurement system.

### III. Results and Discussion

A representative AFM topograph of N-polar GaN films grown in this study is shown in Fig. 1. All samples exhibited a morphology devoid of hexagonal hillocks and a root mean square (RMS) roughness measured over a 90x90 μm$^2$ area below 10 nm. Typical dislocation density was estimated to be ~$10^{10}$ cm$^{-2}$.

Figure 2 shows room temperature hole concentration as a function of the Mg doping concentration. The RT hole concentration exhibited an increase proportional to [Mg] up to a concentration of ~$1.3x10^{19}$ cm$^{-3}$; subsequently, it experienced a sharp drop.

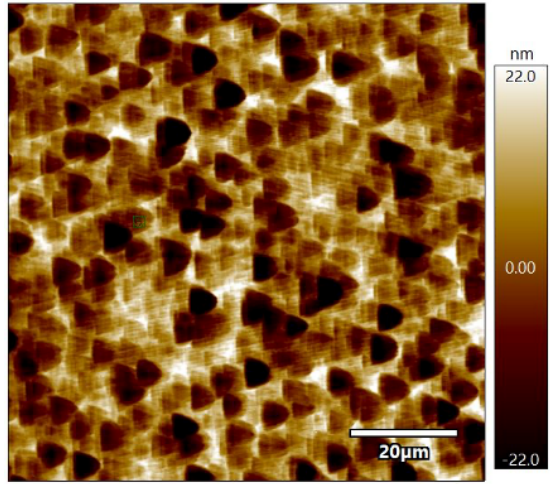

FIG. 1. A representative AFM topograph of a Mg-doped, N-polar GaN film ([Mg]=3x10$^{19}$ cm$^{-3}$).

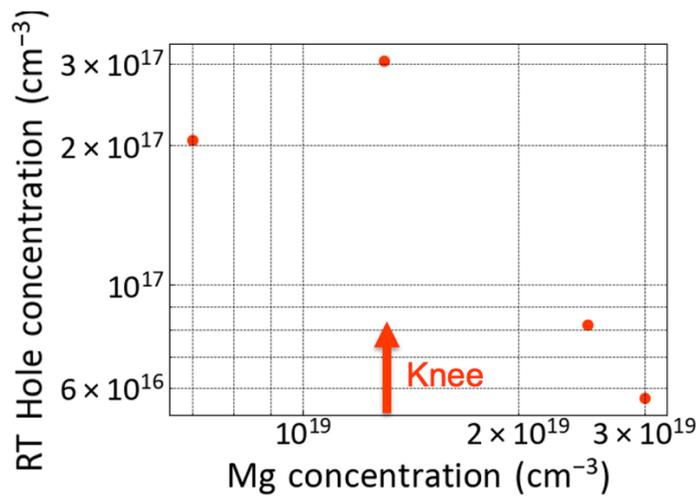

FIG. 2. Room-temperature hole concentration as a function of [Mg].

## A. Charge Balance Model

A sudden drop in hole concentration could have two causes: incorporation of Mg in a neutral state, manifested as a decrease in $N_A$, or an increase in compensation, as a consequence of increased $N_D$. To examine the two possibilities, we quantitatively determined acceptor ($N_A$) and donor ($N_D$) concentrations using a model based on the Fermi Dirac statistics and charge neutrality.

Charge neutrality requires that the concentration of ionized acceptors, $N_A^-(E_F)$, at a particular Fermi level, $E_F$, is equal to the sum of free hole concentration, $p(E_F)$, and the concentration of ionized compensators, $N_D^+$:[34]

$$p(E_F) + N_D^+ = N_A^-. \tag{1}$$

Since all donor-type compensators in p-type materials are ionized, $N_D^+ = N_D$. The concentration of ionized acceptors, $N_A^-(E_F)$, is given by

$$N_A^-(E_F) = \frac{N_A}{1+g\exp\left(\frac{E_A-E_F}{k_BT}\right)}, \tag{2}$$

where $g$ is the degeneracy factor of the valence band, taken to be 4 considering the coincidence of light and heavy hole bands, $k_B$ is the Boltzmann's constant, $E_A$ is the acceptor ionization energy, and $T$ is the absolute temperature. Taking into account the lowering of the ionization energy due to Coulombic screening from free carriers and charge compensators, $E_A$ becomes[22,35]

$$E_A = E_A^0 - b(N_A^-)^{\frac{1}{3}} = 0.22 \text{ eV} - b(p(E_F) + N_D)^{\frac{1}{3}}, \tag{3}$$

where $E_A^0$ is the ionization energy of an isolated Mg acceptor, considered to be 0.22 eV,[22,36] and $b$ is the ionization energy coefficient, determined in this study to be $5.3\times10^{-8}$ eV·cm. The hole concentration as a function of the Fermi level is then[34]

$$p(E_F) = \int_{-\infty}^{0} N_v \frac{2}{\sqrt{\pi}} \frac{\left(\frac{-E}{k_BT}\right)^{\frac{1}{2}}}{1+\exp\left(\frac{E_F-E}{k_BT}\right)} \frac{dE}{k_BT}, \tag{4}$$

where $N_V$ is the effective density of states in the parabolic valence band approximation, and $m^*$ is the effective hole mass, taken to be $2.4m_o$ ($m_o$: electron mass).[37,38]

Using Eqs. (1)-(4) and assuming the Hall scattering factor ($r_H$) to be unity across the studied temperature range (220-440 K),[22,39,40] the acceptor ($N_A$) and donor ($N_D$) concentrations were extracted as follows:

1. The Fermi level, $E_F^{Hall}$, was calculated for each measured hole concentration $p(E_F^{Hall})$ at various temperatures from Eq. (4), using the Newton-Raphson method;

2. Eqs. (2)-(4) were substituted into Eq. (1) to obtain

$$\int_{-\infty}^{0} N_v \frac{2}{\sqrt{\pi}} \frac{\left(\frac{-E}{k_B T}\right)^{\frac{1}{2}}}{1+\exp\left(\frac{E_F-E}{k_B T}\right)} \frac{dE}{k_B T} + N_D = \frac{N_A}{1+g\exp\left(\frac{0.22}{k_B T} - \frac{b}{k_B T}\left(\int_{-\infty}^{0} N_v \frac{2}{\sqrt{\pi}} \frac{\left(\frac{-E}{k_B T}\right)^{\frac{1}{2}}}{1+\exp\left(\frac{E_F-E}{k_B T}\right)} \frac{dE}{k_B T} + N_D\right)^{\frac{1}{3}} - \frac{E_F}{k_B T}\right)} \; ; (5)$$

3. $E_F$ in Eq. (5) was fitted to $E_F^{Hall}$ obtained in step 1 via the least squares method, using $N_A$ and $N_D$ as fitting parameters. This step was enabled by assuming that $r_H = 1$, which is a reasonable assumption for the studied temperature range.[22,39,40]

The obtained results are shown in Figs. 3(a) and (b). Consistent with the RT observations, temperature-dependent hole concentration in Fig. 3(a) shows a similar behavior as a function of [Mg] over the entire studied temperature range: first a proportional increase, and then a sharp drop for [Mg] >1.3x10$^{19}$ cm$^{-3}$. The dots in Fig. 3(a) are experimental data points while the solid lines correspond to $p(E_F)$ obtained from Eq. (4). Figure 3(b) shows the extracted $N_A$ and $N_D$ values for different Mg concentrations. Interestingly, the $N_D$ values remained low, as compared to [Mg], for all doping levels and were more or less flat on the order of mid-10$^{17}$ cm$^{-3}$. In contrast, $N_A$ closely matched [Mg] up to ~1.3x10$^{19}$ cm$^{-3}$, followed by a significant drop for [Mg] >1.3x10$^{19}$ cm$^{-3}$. The values of the extracted Mg ionization energy ($E_A$) at 300 K ranged from 160-180 meV, agreeing well with the typically reported values for Ga- and N-polar Mg:GaN.[8–10] The compensation ratio ($N_D/N_A$) was <10% at the knee point and increased up to ~30% for higher [Mg], as manifested by a practically constant $N_D$ and a reduced $N_A$. The relatively low $N_D$ rules out the dominance of charged compensating defects such as nitrogen-vacancies ($V_N^{+3}$), charged $V_N$-complexes (($V_N$-nMg$_{Ga}$)$^{3-n}$, where n = 1 or 2)[17–19] and Mg interstitials (Mg$_{inter}^{+2}$).[41] Additionally, [O] measured by SIMS remained constant at ~$4\times10^{17}$ cm$^{-3}$ and was independent of [Mg], as illustrated by the dashed line in Fig. 3(b). This indicates that the majority of $N_D$ consisted of $O_N^{+1}$ rather than $(V_N-nMg_{Ga})^{3-n}$ where n = 0, 1, and 2, consistent with observations for n-type N-polar GaN, where $O_N^{+1}$ is the dominant donor.[12,42,43]

All of the above findings indicate that the knee behavior in hole concentration stems from a drastic reduction in $N_A$, suggesting that for high [Mg] a significant fraction of the Mg atoms is incorporated in an electrically neutral configuration. The possible candidates for this neutral state include $(V_N-3Mg_{Ga})^0$ complexes or PIDs in the form of $Mg_3N_2$ precipitates.[17,21,23] Our initial TEM studies along with other published work[25,26] suggest that the amount of Mg segregated at PIDs is insufficient to explain this knee behavior. Interestingly, MOCVD-grown p-type Ga-polar GaN exhibits a knee behavior at a higher [Mg] of ~$3\times10^{19}$ cm$^{-3}$,[17–22] differing from the knee onset at ~$1.3\times10^{19}$ cm$^{-3}$ observed in the p-type N-polar GaN. This difference implies that the defect responsible for the knee behavior is surface/polarity-dependent and forms at the surface during growth. Given the possibilities discussed above, the $(V_N-3Mg_{Ga})^0$ complex—surface/polarity-dependent—is a more likely cause for the knee behavior than the PIDs—bulk-dependent.[23]

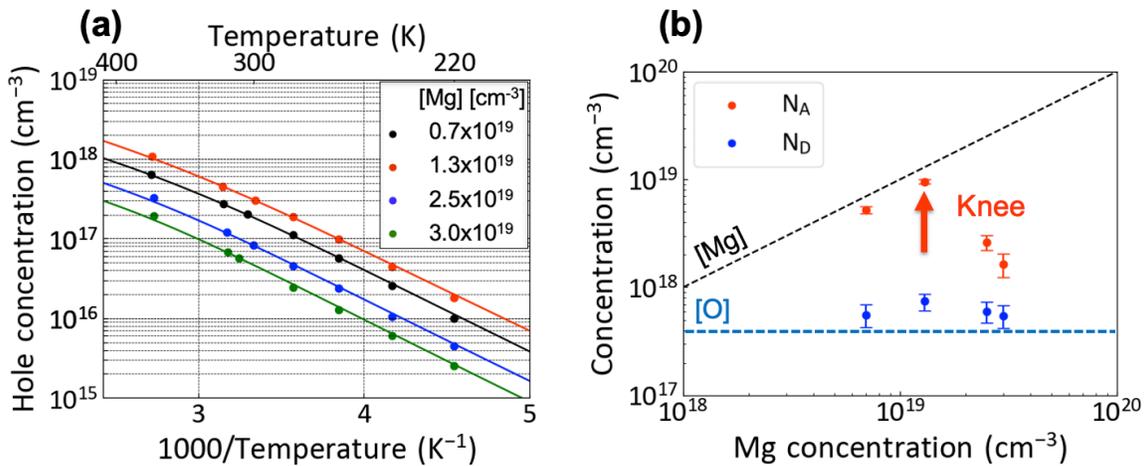

FIG. 3. (a) Temperature dependent hole concentration (dots experimental, solid fit) for various Mg doping concentrations. (b) Extracted acceptor ($N_A$) and donor ($N_D$) concentrations as a function of [Mg]. Error bars represent the error in $N_A$ and $N_D$ obtained from the charge balance model based on the least squares fitting, assuming $r_H = 1$. SIMS-measured [O] remained constant at ~4x10$^{17}$ cm$^{-3}$ and was practically independent of [Mg].

**B. Semi-Empirical Model Based on the Grand Canonical Formalism**

To gain quantitative insight into the ($V_N$-3Mg$_{Ga}$)$^0$ complex, we developed a semi-empirical model based on the grand canonical formalism that relates the propensity of incorporating a particular point defect during a growth process to its formation energy:[43]

$$E^f(X^q) = E_{\text{ref}}(X^q) + \sum_j n_j \mu_j + q[E_F + E_V], \qquad (6)$$

where $E^f(X^q)$ is the formation energy for the point defect $X$ in a charge state $q$, $E_{\text{ref}}(X^q)$ is the free energy of the crystal with a single defect referenced to the free energy of an ideal crystal, $n_j$ is the number of atoms of $j^{th}$ type exchanged with the reservoir to form the point defect $X$, $\mu_j$ is the associated chemical potential, and $E_F$ is the Fermi level referenced to the valence band maximum. Thus, the formation energy of the neutral ($V_N$-3Mg$_{Ga}$)$^0$ complex can be written as

$$E^f((V_N - 3Mg_{Ga})^0) = E_{\text{ref}}((V_N - 3Mg_{Ga})^0) - 3\mu_{Mg} + \mu_N, \qquad (7)$$

where $\mu_N$ and $\mu_{Mg}$ are the chemical potentials of nitrogen and magnesium, respectively. It is assumed that the Mg dopant incorporation on the Ga site is spontaneous, i.e., $E^f(Mg_{Ga}) < 0$.[44]

The only variation in this study is the Mg doping concentration and with that the Mg chemical potential, $\mu_{Mg}$. Therefore, the change in the formation energy relative to a reference growth state depends only on the change in $\mu_{Mg}$. This reduces Eq. (7) to:

$$\Delta E^f((V_N - 3Mg_{Ga})^0) = -3\Delta\mu_{Mg}, \qquad (8)$$

The corresponding concentration of the neutral complex is then given by[27,43]

$$[(V_N - 3Mg_{Ga})^0]_{exp} = [(V_N - 3Mg_{Ga})^0]_{ref} \exp\left(\frac{3\Delta\mu_{Mg}}{k_B T}\right), \tag{9}$$

where $[(V_N - 3Mg_{Ga})^0]_{exp}$ and $[(V_N - 3Mg_{Ga})^0]_{ref}$ are the concentrations of $(V_N\text{-}3Mg_{Ga})^0$ for the experimental and reference growth states, respectively.

Central to this semi-empirical model is the relationship between $\Delta\mu_{Mg}$ and the change of the concentration of the neutral $(V_N\text{-}3Mg_{Ga})^0$ complex, where the Mg chemical potential at the surface with respect to an arbitrarily chosen reference state is given by[45]

$$\mu_{Mg} = \mu_{Mg}^o + k_B T \ln\left(\frac{p_{Mg}^*}{p_{Mg}^o}\right), \tag{10}$$

where $p_{Mg}^*$ is the input Mg partial pressure, and $\mu_{Mg}^o$ and $p_{Mg}^o$ are the Mg chemical potential and partial pressure of a reference state, respectively. Since $E^f(Mg_{Ga}) < 0$, the Mg concentration is proportional to $p_{Mg}^*$ and inversely proportional to the growth rate ($v_{GR}$), i.e., $[Mg] \propto \frac{p_{Mg}^*}{v_{GR}}$.[46] Thus, the relative change in $\mu_{Mg}$ with respect to a reference growth state can be expressed as:

$$\Delta\mu_{Mg} = k_B T \ln\left(\frac{p_{Mg}^*}{\{p_{Mg}^*\}_{ref}}\right) = k_B T \ln\left(\frac{[Mg] \, v_{GR}}{[Mg]_{ref} \, (v_{GR})_{ref}}\right), \tag{11}$$

Substituting Eq. (11) into Eq. (9) yields:

$$[(V_N - 3Mg_{Ga})^0]_{exp} = [(V_N - 3Mg_{Ga})^0]_{ref} \frac{v_{GR}^3}{[Mg]_{ref}^3 (v_{GR})_{ref}^3}[Mg]^3 = A[Mg]^3, \tag{12}$$

where $A$ is a fitting parameter in units of $[cm^6]$. The number of active Mg acceptors ($N_A$) can then be calculated as the difference between the total $[Mg]$ and 3x complex concentration, since each complex inactivates three Mg atoms:

$$N_A = [Mg] - 3[(V_N - 3Mg_{Ga})^0]_{exp} = [Mg] - 3A[Mg]^3. \tag{13}$$

The $N_A$ values obtained from Eq. (13) are plotted in Fig. 4 along with the experimental data points. As can be seen, the above semi-empirical model describes the observed "knee" behavior qualitatively and quantitatively and can be used to estimate $N_A$ values for specific [Mg] and growth conditions.

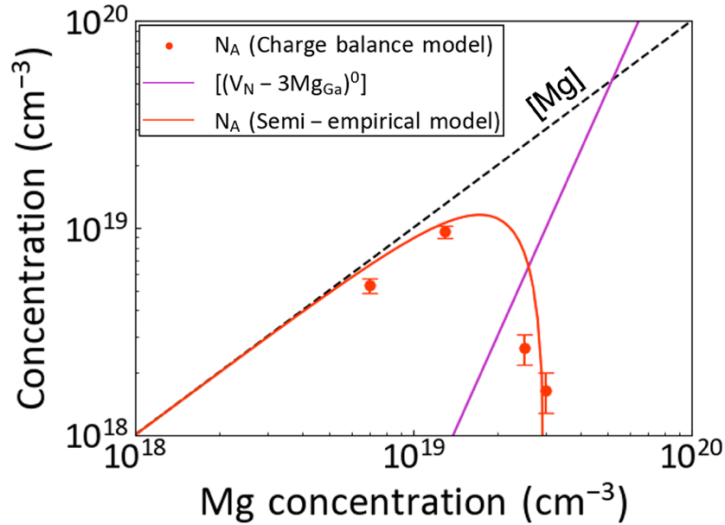

FIG. 4. Mg acceptor concentration ($N_A$) as a function of [Mg]: dots represent values obtained from the charge balance model while the solid red line represents $N_A$ calculated using the semi-empirical model, when $A=3.7 \times 10^{-40}$. The purple line shows the corresponding concentration of $(V_N-3Mg_{Ga})^0$.

## IV. Conclusion

In summary, this study identified the neutral $(V_N-3Mg_{Ga})^0$ complex as the primary cause of the knee behavior in Mg-doped, N-polar GaN. The self-compensation mechanism was analyzed using a charge balance model in conjunction with a semi-empirical model based on the grand canonical formalism. The charge balance model revealed that the knee behavior was a consequence of a sharp drop in $N_A$ rather than a surge in $N_D$, indicating that a significant fraction of Mg dopants was incorporated in an electrically neutral configuration, while the semi-empirical model based on the grand canonical formalism provided good quantitative agreement with the experimental data. A comparison between this study for the N-polar GaN and Ga-polar GaN studies from literature suggests that the defect responsible for the knee behavior in both cases is the surface/polarity-dependent $(V_N\text{-}3Mg_{Ga})^0$ complex rather than pyramidal inversion domains (PIDs).

**Acknowledgements**

The authors gratefully acknowledge funding in part from AFOSR (FA9550-17-1-0225, FA9550-19-1-0114), DoD ARPA-E (DE-AR0001493), and NSF (ECCS-1508854, ECCS-1916800, ECCS-1653383).